\documentclass[sigconf]{acmart}

\renewcommand\footnotetextcopyrightpermission[1]{} 
\usepackage{array}
\usepackage{CJKutf8}
\usepackage[linesnumbered,ruled,vlined]{algorithm2e}

\usepackage{fancyhdr}
\usepackage{multirow}
\usepackage{graphicx}
\usepackage{subcaption}

\AtBeginDocument{%
  \providecommand\BibTeX{{%
    \normalfont B\kern-0.5em{\scshape i\kern-0.25em b}\kern-0.8em\TeX}}}

\setcopyright{acmcopyright}
\copyrightyear{2018}
\acmYear{2018}
\acmDOI{XXXXXXX.XXXXXXX}

\acmPrice{15.00}
\acmISBN{978-1-4503-XXXX-X/18/06}

\begin{document}
\begin{CJK}{UTF8}{gbsn}
\title{EnvGuard: Guaranteeing Environment-Centric Safety and Security Properties in Web of Things}



\author{Bingkun Sun, Liwei Shen, Jialin Ren, Zhen Dong, Siao Wang, Xin Peng}
\affiliation{%
  \country{Department of Computer Science, Fudan University, China}
}
\email{21110240011, 22212010029, 22110240039@m.fudan.edu.cn}
\email{shenliwei, zhendong, pengxin@fudan.edu.cn}
\begin{abstract}
Web of Things (WoT) technology facilitates the standardized integration of IoT devices ubiquitously deployed in daily environments, promoting diverse WoT applications to automatically sense and regulate the environment. 
In WoT environment, heterogeneous applications, user activities, and environment changes collectively influence device behaviors, posing risks of unexpected violations of safety and security properties.
Existing work on violation identification primarily focuses on the analysis of automated applications, lacking consideration of the intricate interactions in the environment. Moreover, users’ intention for violation resolving strategy is much less investigated.
To address these limitations, we introduce EnvGuard, an environment-centric approach for property customizing, violation identification and resolution execution in WoT environment. 
We evaluated EnvGuard in two typical WoT environments. By conducting user studies and analyzing collected real-world environment data, we assess the performance of EnvGuard, and construct a dataset from the collected data to support environment-level violation identification.
The results demonstrate the superiority of EnvGuard compared to previous state-of-the-art work, and confirm its usability, feasibility and runtime efficiency.

\end{abstract}








\maketitle

\section{Introduction}
The rapid development of Internet of Things (IoT) has accelerated the deployment of smart devices in daily environments. Building on this, Web of Things (WoT) standardizes the integration of fragmented IoT device as perceptual event and executable action service using web technologies~\cite{Sun2023SCTAP, Zyrianoff2022SeamlessIO}, enhancing device interoperability~\cite{9847610,10.1145/3137133.3141466} and facilitating the development of various WoT applications for automated sensing and regulation of environments~\cite{Guinard2011FromTI}.

Although the WoT paradigm brings great convenience for daily life, safety and security concerns attract great attention~\cite{Pradeep2021AutomatingCD, Huang2021ConflictDI, Chaki2020ACD, Chaki2020FinegrainedCD, Wang2019ChartingTA, Shah2019ConflictDI}. It is crucial to guarantee the device behavior complies with the desired environment \textbf{properties}, i.e., the safety and security constraints that should always be satisfied by the WoT environment.
For example, a common safety concern is the improper use of electrical devices like air conditioners or lights, posing threats to people's health and the electrical safety standards. Another security concern is the risk of property loss and privacy breach due to windows and curtains remaining open when the house is empty.

To this end, existing research contributes to identifying property violations by analyzing the impact of WoT applications on the state of devices, and resolves them by intercepting or disabling applications~\cite{Soteria, IoTSan, SafeChain, IoTIE, HomeGuard, IoTCom, IoTSafe, AutoTap}.
However, \textbf{limitations} still remain in many aspects.

\textbf{Property violations are not only caused by applications}, but also by user activities and environment changes which current application-centric methods fail to identify.
For example, a user manually opens the meeting room window on a sunny day and leaves, and subsequently the weather changes to rain, posing a risk of rainwater damage to indoor items.

\textbf{The interplay between environment context and the context-sensitive device services} implies that current methods lack analysis of the intricate interrelation.
For example, opening and closing a window can have varying effects on indoor temperature, humidity, and air quality under different outdoor conditions. Identifying violations only based on the application-related device state can lead to omissions or misjudgments.

\textbf{The definition of property} fails to meet the customization requirements when it is predefined and built-in, 
while accurately specifying properties is challenging for non-technical users. 
For example, to specify properties related to the environment temperature, users must consider all devices potentially affecting temperature (e.g., air conditioner, window, and heater) and the related environment context (e.g., weather, outdoor temperature). 

\textbf{Environment properties span both spatial and temporal aspects}, while existing methods primarily address spatial environment context, neglecting the temporal trace of device services. 
For instance, ensuring electrical safety by detecting prolonged lighting in empty rooms requires a temporal analysis of human detection events and light-switch actions.

\textbf{The intercepting or revoking resolving strategy may be inconsistent with user intention}, as the violation actions possibly contain user requirements. For example, if a user opens a window while raining, closing it as a resolution might be contrary to the ventilation intent.

\begin{figure}[ht]
	\centering
	\includegraphics[width = 0.43\textwidth]{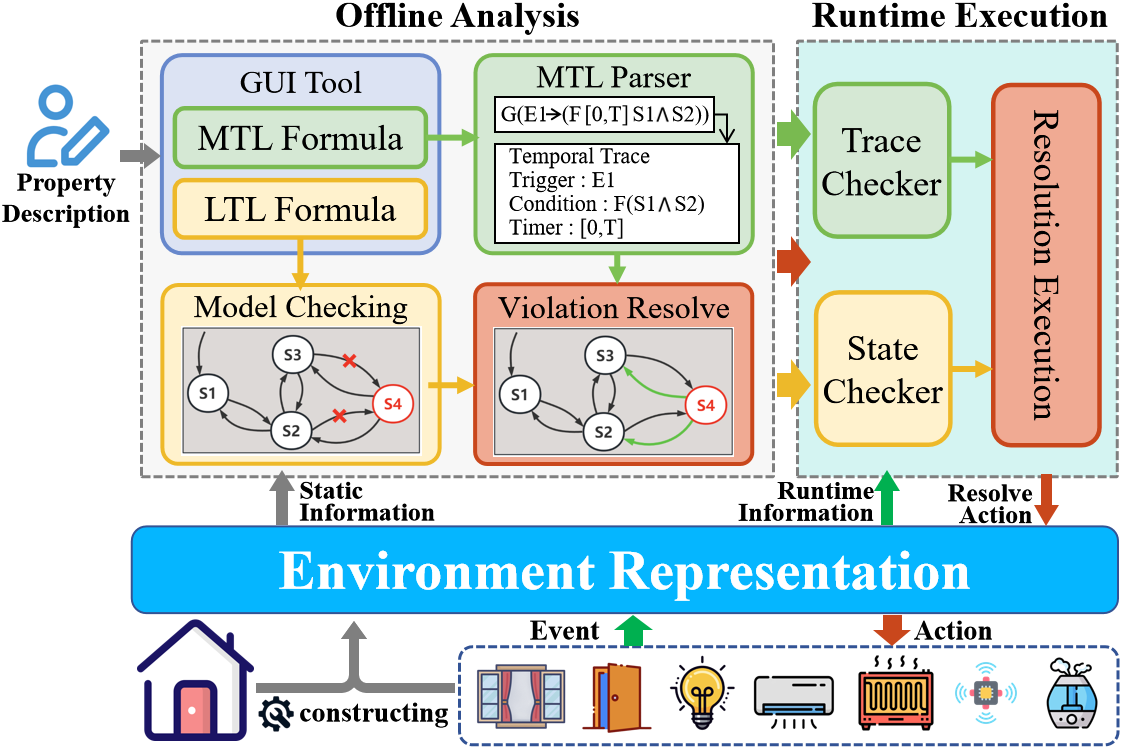}
        \vspace{-10pt}
    	\caption{An approach overview of EnvGuard}
        \Description{The EnvGuard includes an offline phase to identify the violation and find a repair strategy, and a runtime engine to check and repair the violation behavior}
    \label{Approach-Overvire}
\end{figure}

To address these limitations, we present EnvGuard, an environment-centric approach that offers expressive descriptions and comprehensive guarantees for properties, as shown in Figure \ref{Approach-Overvire}

EnvGuard first defines a conceptual schema to model the interplay between environment context and context-sensitive device services, and automatically generates a representation of a WoT environment based on device and space information.
Then, a GUI tool is provided to assist users in customizing spatial and temporal environment properties by instantiating property description templates, which can be translated into linear temporal logic (LTL) and metric temporal logic (MTL) formulas.
During offline phase, EnvGuard separately analyzes the LTL describing the spatial state properties and the MTL describing the temporal trajectory properties by adopting a hybrid model-checking technique, identifying potential violations and their possible resolution actions. 
According to the result of the user study on user preference for resolving different types of violations, three optional resolving strategies are proposed to ensure the consistency of resolution action and user's intention.
At runtime, EnvGuard checks spatial state and temporal trace property in an event-driven manner and, based on the resolving strategy, selects non-conflicting actions to resolve identified violations without introducing new ones.

To evaluate EnvGuard, we implement it as a prototype and conduct two representative real-world case studies including a smart office and a smart home.
The evaluation demonstrates that EnvGuard can facilitate users in constructing spatio-temporal environment properties, accurately identifies all property violations, and generates resolution actions aligned with user intentions while ensuring real-time performance.


We summarize our main \textbf{contributions} as follows:

\textbf{Violation Identification}. We introduce a schema to model the interplay between environment context and devices, enabling comprehensive identification of spatio-temporal property violations caused by various entities (including application, user activities, and environment changes) through a hybrid model-checking technique.

\textbf{Resolving Strategy}. We investigate user preferences for resolution actions, offering selectable resolving strategies for different types of properties to ensure alignment with user intentions.

\textbf{Property Customizing}. We provide multiple property description templates and corresponding GUI tool to assist users in customizing environment properties.

\textbf{Practical Results}. We conduct two real-world case studies to demonstrate EnvGuard's performance, and construct the collected environment data as the first dataset\cite{EnvGuard} which records events and actions in the environment and manually labels 20 violation types.

\section{Environment Modeling}
In this section, we first present a schema to model the interplay between environment context and device service, and introduce its automatic construction based on device and space information.

\vspace{-9pt}
\subsection{Conceptual Schema of Environment}
\begin{figure}[htbp]
	\centering
	\includegraphics[width = 0.38\textwidth]{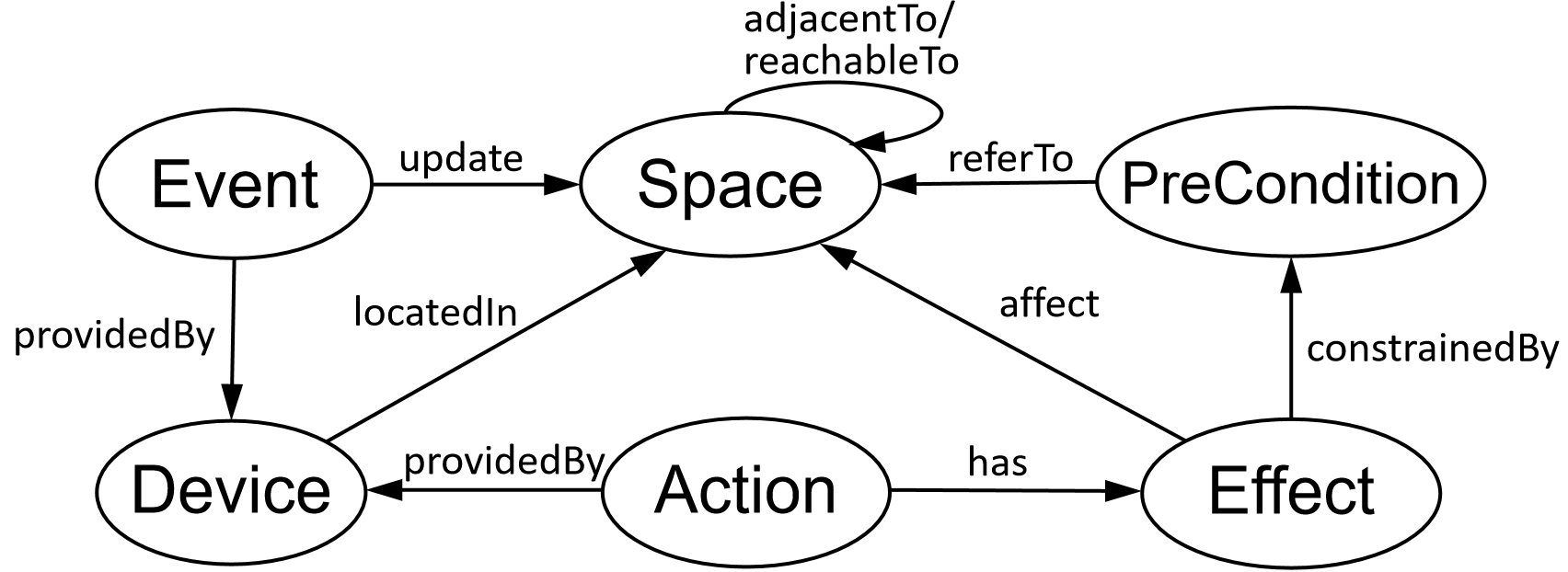}
        \vspace{-10pt}
    	\caption{Conceptual schema of the environment}
        \vspace{-3pt}
        \Description{The conceptual schema includes "Space", "Device", "Event", "Action", "Effect", and "PreCondition"}
    \label{Schema}
\end{figure}

The conceptual schema is illustrated in Figure~\ref{Schema}, describing the concepts and the relationship of the spaces, the events and actions provided by devices, and the effects that actions can produce on the environment under the specific preconditions. We further formalize each concept in the conceptual schema.

\emph{\textbf{Space}} describes the space within the environment, which is formally defined as a tuple of \textit{<id, type, Attributes, States>}. id is a unique identifier of the space and type is the category of the space (e.g., meeting room, corridor, etc.). Attributes are a set of space attributes defined as a map $\{(a_1:v_1), ..., (a_n:v_n)\}$ where each $a_i$ is the attribute name (e.g., area) and $v_i$ is the corresponding value (e.g., 50 $m^2$). And States is a set of variables of the space defined as a map $\{(s_1:v_1), ..., (s_n:v_n)\}$ where each $s_i$ is the state name (e.g., temperature) and $v_i$ is the corresponding value. The continuous state values are discretized (e.g., discretize the temperature into high, medium, and low). The topological structure of different spaces is described by the "adjacentTo" and "reachableTo" relationship.


\textbf{Device} describes the deployed device, and is represented as a tuple of \textit{<id, type, state>}, indicating the unique identifier, category (e.g., window, light), and current state (e.g., on, off). The "locatedIn" relationship specifies the space where a device is deployed.

\emph{\textbf{Event}} is a device service that describes changes in the environment state and is provided by sensors.
Event is defined as a tuple of \textit{<event\_type,  location, target\_state, payload, Topic>}. event\_type indicates the category of the event (e.g., Temperature\_Change), location indicates the space where the event occurs, target\_state specifies the space state impacted by the event (e.g., temperature), and payload is the current value of the changed state (e.g., high). Topic is defined as a map $\{(id_1:t_1), ..., (id_n:t_n)\}$ where each $id_i$ indicates the id of device that can provide the event, and each $t_i$ indicates the message queue topic for subscribing to the event provided by the device. The "providedBy" relationship specifies the device providing the event, and the "update" relationship specifies the target space where the state change has occurred.

\emph{\textbf{Action}} is the other device service that describes the executable function of actuators, which is defined as a tuple of \textit{<action\_type, location, target\_state, payload, URL>}. 
action\_type categorizes the action (e.g., AC\_TurnOn), location specifies the execution site, target\_state indicates the device whose state is impacted by the action, and the payload is the value of the current device state.
URL is a map $\{(id_1:url_1), ..., (id_n:url_n)\}$ where each $id_i$ is a device ID offering the action, and each $url_i$ is the API address for the action service.
The "providedBy" relationship specifies the device performing the action, and the "has" relationship specifies the action's impact on environment contexts.


\emph{\textbf{Effect}} describes the impact of an action on the environment state, which is defined as a pair of \textit{<state\_name, effect\_type>}. state\_name specifies the space state impacted by the action (e.g., temperature), and effect\_type specifies the effect of the action on the target\_state (e.g., increase, decrease, etc.). The "affect" relationship indicates the space impacted by this effect, and the "constrainedBy" relationship specifies the constraint condition.

\emph{\textbf{PreCondition}} indicates the prerequisite environment conditions that need to be satisfied for the effect to take place. PreCondition is defined as a tuple of \textit{<state\_name, operator, target\_value>}, where state\_name specifies the space state that needs to be satisfied, operator specifies the comparison operator (e.g., >, <, etc.), and target\_value specifies the value that needs to be satisfied. Specifically, the target\_value can be set either by a specific value or referring to a space state. For example, the target\_value can be set to high or refer to the temperature of the adjacent space. The "referTo" relationship specifies the space that the state\_name and target\_value refer to.

\subsection{Environment Representation Construction}
We introduce a device description and space information format, allowing WoT developers and device vendors to provide standardized device and space data. This facilitates the automatic creation of the environment representation based on our conceptual schema.
Compared to existing device description formats such as Azure IoT\cite{Azure}, W3C ThingDescription\cite{TD}, and AWS IoT \cite{AWS}, 
our format builds on basic concepts like device state and function, while introducing new concept of effect and precondition to describe the interplay between device service and environment context.
Additionally, the space information format is provided to describe spatial attributes, states, and topological structures.

\begin{figure}[htbp]
	\centering
	\includegraphics[width = 0.42\textwidth]{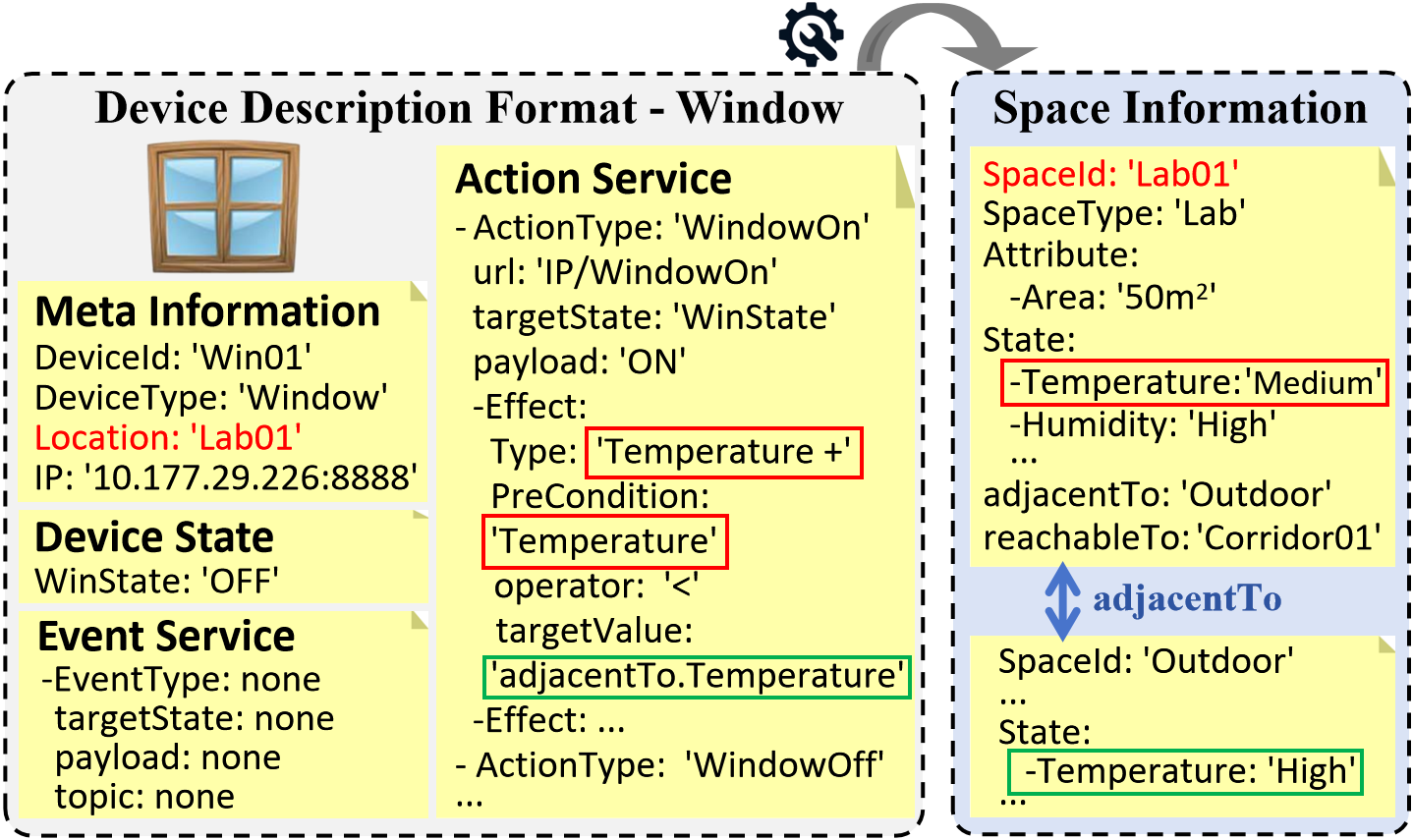}
        \vspace{-10pt}
    	\caption{An example of associating device description format with space information format}
        \Description{The device description format includes device information, device state, event service, and action service}
    \label{Construction}
\end{figure}

Figure \ref{Construction} illustrates the space information format and the device description format along with their relationships through an example. 
The space information format describes the id, type, properties, states, and relationships with other spaces. By instantiating all space information to space instances and connecting them based on the position relationships (e.g., the lab and outdoor instances are connected by the "adjacentTo" relationship), the representation of the space concept can be completed.

In the device description format, MetaInformation and DeviceState respectively depict the basic information and work state of the device, which is utilized to generate a device instance, and the instance is connected to the event and action instances provided by the device. 
EventService outlines all events offered by the device. 
Equivalent events from different devices are instantiated only once by the first device, recording all corresponding device IDs and subscription topics. 
ActionService describes all actions provided by the device, detailing the consequent effects under diverse preconditions. 
Equivalent action services with identical effects and preconditions from different devices are also instantiated into a single action instance, recording all corresponding device IDs and API addresses, and connecting to the corresponding effect and precondition instances. 

All instances generated by the device description format can be automatically associated with space instances based on the device's location information (e.g., the window instance is connected to the Lab01 instance through the "locatedIn" relationship), completing the construction of environment representation. 

In implementation, we use a graph database (i.e., \emph{Neo4J}) to create and maintain the instances and relationships of the environment representation, where instances correspond to nodes, and relationships between instances correspond to directed edges between nodes. In offline phase, environment representation offers static interplay information between device services and space states. At runtime, the representation updates corresponding space and device states according to events and actions. By querying the graph database, environment  information can be accessible.

\section{Property Description}

EnvGuard employs the widely adopted template approach\cite{Autili2015AligningQR, Dou2014AMA, Lumpe2011PSPWizardMD} to support end-users in customizing desired environment properties. 
The users specify environment properties by selecting templates and instantiating them with specific values for placeholders in the templates.
These templates can be translated into formal language such as linear temporal logic(LTL) and metric temporal logic(MTL), which are widely used for property description and system verification~\cite{Clarke1986AutomaticVO}. 
An LTL formula consists of atomic propositions (basic statements that cannot be further decomposed, i.e., the state value of space and device), logical operators (such as and $\land$, or $\lor$, and implies $\rightarrow$), and temporal operators (such as globally G and finally F). MTL extends LTL by adding time-related metrics to specify time bounds for temporal operators.

For the commonly desired property types in WoT environment~\cite{A3ID, Ibrhim2021ACC, AutoTap, Sun2023SCTAP}, EnvGuard provides eight templates to support users in describing violations by assigning states, events, or actions to placeholders. 
Table \ref{tab:template} lists the templates, where LTL-type templates describe violated spatial state, 
and MTL-type templates describe temporal state trace violation.

\begin{table*}[h]
\renewcommand{\arraystretch}{0.5}
\caption{Property description templates and the corresponding LTL/MTL formulas. The superscript "\( ^+ \)" indicates that multiple attributes can be specified.}
\vspace{-8.5pt}
\label{tab:template}
\resizebox{\linewidth}{!}{%
\begin{tabular}{@{}cllll@{}}
\toprule
Type & Id & Property Description Template & {LTL/MTL Formula} \\ 
\midrule
\multirow{8}{*}{\raisebox{-1.05cm}{LTL}} 
& \multirow{2}{*}{\#1} 
& {[}\(\textit{effect}^+\){]} should {[}\textit{always}{]} be executed simultaneously 
& \( G(\textit{effect}_1 \leftrightarrow \ldots \leftrightarrow \textit{effect}_n) \) \\
& & {[}\(\textit{effect}^+\){]} should {[}\textit{never}{]} be executed simultaneously 
& \( G\neg(\textit{effect}_1 \wedge \ldots \wedge \textit{effect}_n) \) \\
\cmidrule(l){2-4} 
& \multirow{2}{*}{\#2} 
& {[}\(\textit{effect}\){]} should {[}\textit{always}{]} be executed under {[}\(\textit{state}^+\){]} 
& \( G((\textit{state}_1 \wedge \ldots \wedge \textit{state}_n) \rightarrow \textit{effect}) \) \\
& & {[}\(\textit{effect}\){]} should {[}\textit{never}{]} be executed under {[}\(\textit{state}^+\){]} 
& \( G((\textit{state}_1 \wedge \ldots \wedge \textit{state}_n) \rightarrow \neg\textit{effect}) \) \\
\cmidrule(l){2-4} 
& \multirow{2}{*}{\#3} 
& {[}\(\textit{state}^+\){]} should {[}\textit{always}{]} happen simultaneously 
& \( G(\textit{state}_1 \leftrightarrow \ldots \leftrightarrow \textit{state}_n) \) \\
& & {[}\(\textit{state}^+\){]} should {[}\textit{never}{]} happen simultaneously 
& \( G\neg(\textit{state}_1 \wedge \ldots \wedge \textit{state}_n) \) \\
\cmidrule(l){2-4} 
& \multirow{2}{*}{\#4} 
& {[}\(\textit{event/action}\){]} should {[}\textit{only}{]} happen under {[}\(\textit{state}^+\){]} 
& \( G(\textit{event/action} \rightarrow (\textit{state}_1 \wedge \ldots \wedge \textit{state}_n)) \) \\
& & {[}\(\textit{event/action}\){]} should {[}\textit{never}{]} happen under {[}\(\textit{state}^+\){]} 
& \( G(\textit{event/action} \rightarrow \neg(\textit{state}_1 \wedge \ldots \wedge \textit{state}_n)) \) \\
\midrule
\multirow{8}{*}{\raisebox{-1.0cm}{MTL}} 
& \multirow{2}{*}{\#5} 
& {[}\(\textit{effect}\){]} should last {[}\textit{more}{]} than {[}\(\textit{time}\){]} 
& \( G(\textit{effect} \rightarrow (G[0, \textit{time}] \textit{effect})) \) \\
& & {[}\(\textit{effect}\){]} should last {[}\textit{less}{]} than {[}\(\textit{time}\){]} 
& \( G(\textit{effect} \rightarrow (F[0, \textit{time}] \neg \textit{effect})) \) \\
\cmidrule(l){2-4} 
& \multirow{2}{*}{\#6} 
& {[}\(\textit{effect}\){]} should {[}\textit{always}{]} be executed within {[}\(\textit{time}\){]} of entering {[}\(\textit{state}^+\){]} 
& \( G((\textit{state}_1 \wedge \ldots \wedge \textit{state}_n) \rightarrow (F[0, \textit{time}] \textit{effect})) \) \\
& & {[}\(\textit{effect}\){]} should {[}\textit{never}{]} be executed within {[}\(\textit{time}\){]} of entering {[}\(\textit{state}^+\){]} 
& \( G((\textit{state}_1 \wedge \ldots \wedge \textit{state}_n) \rightarrow (G[0, \textit{time}] \neg \textit{effect})) \) \\
\cmidrule(l){2-4} 
& \multirow{2}{*}{\#7} 
& {[}\(\textit{state}^+\){]} should {[}\textit{always}{]} happen within {[}\(\textit{time}\){]} when the {[}\(\textit{event/action}\){]} occurs 
& \( G(\textit{event/action} \rightarrow (F[0, \textit{time}] (\textit{state}_1 \wedge \ldots \wedge \textit{state}_n))) \) \\
& & {[}\(\textit{state}^+\){]} should {[}\textit{never}{]} happen within {[}\(\textit{time}\){]} when the {[}\(\textit{event/action}\){]} occurs 
& \( G(\textit{event/action} \rightarrow (G[0, \textit{time}] \neg (\textit{state}_1 \wedge \ldots \wedge \textit{state}_n))) \) \\
\cmidrule(l){2-4} 
& \multirow{2}{*}{\#8} 
& {[}\(\textit{state}^+\){]} should last {[}\textit{more}{]} than {[}\(\textit{time}\){]} 
& \( G((\textit{state}_1 \wedge \ldots \wedge \textit{state}_n) \rightarrow (G[0, \textit{time}] (\textit{state}_1 \wedge \ldots \wedge \textit{state}_n))) \) \\
& & {[}\(\textit{state}^+\){]} should last {[}\textit{less}{]} than {[}\(\textit{time}\){]} 
& \( G((\textit{state}_1 \wedge \ldots \wedge \textit{state}_n) \rightarrow (F[0, \textit{time}] \neg (\textit{state}_1 \wedge \ldots \wedge \textit{state}_n))) \) \\
\bottomrule
\end{tabular}%
}
\end{table*}

\begin{figure}[ht]
	\centering
	\includegraphics[width = 0.34\textwidth]{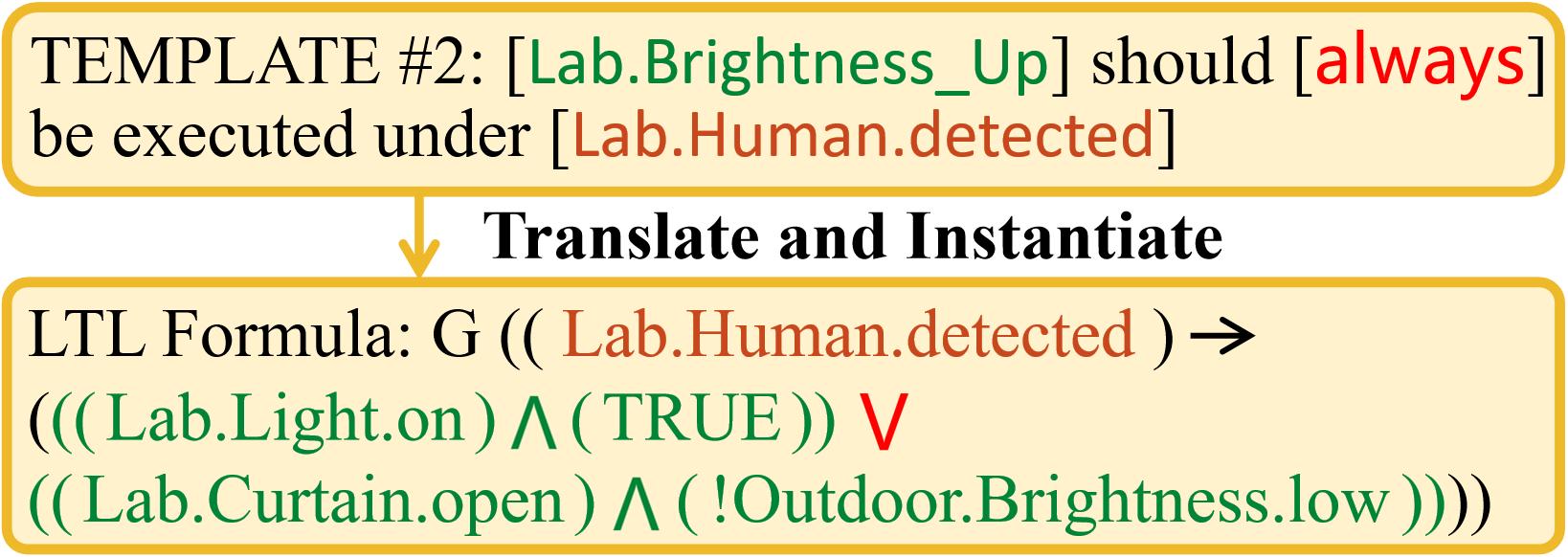}
        \vspace{-10pt}
	\caption{An example of abstract template instantiation}
        \Description{the abstract effect "Lab.Brightness\_Up" is instantiated to the corresponding states of the laboratory light and curtain as well as the environment context}
    \label{AbstractTemplate}
\end{figure}

Although the template approach enables users to customize system properties, 
understanding the interplay between device services and environment context remains complex, which brings challenges to configuring the template accurately.
To alleviate the complexity, environment-centric property description templates are further proposed (template \#1, \#2, \#5, and \#6) to enable user 
to describe device behaviors by specifying the effects of the device being executed in the environment. 
By querying actions with the specified effect as well as the corresponding preconditions in the environment representation, the effect can be instantiated into device state and environment context.
Figure \ref{AbstractTemplate} demonstrates an example of brightness management where the abstract effect "Lab.Brightness\_Up" is instantiated to the corresponding states of the laboratory light and curtain as well as the environment context.

Algorithm \ref{algo-effect} outlines the detailed process of instantiating abstract effects based on the environment representation.
By querying actions with effects and their corresponding device states, and considering environment context that meets the precondition, abstract effects can be mapped to device and environment states. 

\begin{algorithm}[t]
\SetAlgoLined
\caption{Abstract effect instantiation}\label{algo-effect}
\KwIn{effect, environment representation}
\KwOut{instantiatedEffect}

Find all $\textbf{Action}$ with effect in environment representation\;

formulaList $\gets []$\;
\For{action in Action}{
    deviceState $\gets$ findAllState(action, "providedBy")\;
    deviceState $\gets$ deviceState.join("or")\;
    PreCondition $\gets$ findCondition(action, "has", effect)\;
    \eIf{PreCondition not null}{
        Find all spaceState that meet the PreCondition\;
        formulaList.append(deviceState + "and" + spaceState.join("or"))\;
    }{
        formulaList.append(deviceState + "and" + "TRUE")\;
    }
}
\textbf{Return} instantiatedEffect $\gets$ formulaList.join("or")\;
\end{algorithm}

To assist the user in specifying properties, EnvGuard provides a visualization tool implemented by Vue.js to automatically acquire environment information and complete template configurations through mouse clicks. The tool's UI design is similar to the popular smart home automation platforms such as IFTTT\cite{IFTTT} and Zapier\cite{Zapier}, which makes it more accessible to non-technical users.
Users can easily customize properties by selecting a template and instantiating placeholders with environment information from a drop-down list. The user interface can be seen online \cite{EnvGuard}.

\section{Violation Identification And Resolving}
In this section, we illustrate how EnvGuard identifies violations, and the rationale behind the technology is further discussed.
Then three optional resolving strategies are presented to resolve different types of violations.

\subsection{Violation Identification}

\subsubsection{Spatial State Violation Identification}\label{Spatial-State}
EnvGuard utilizes an offline model checking method to identify the presence of states that violate the LTL formula and the transitions that lead to the violated state~\cite{Merz2000ModelC}. 
A typical model checking method takes three steps: 
(1) Model the environment as a B\"{u}chi Automaton \( \mathbb{A}_E \), 
a specialized finite state machine for infinite input sequences~\cite{Sickert2016LimitDeterministicBA}, enabling comprehensive exploration and detection of potential violations.
In this step, \( \mathbb{A}_E \) does not contain any accepting state,
which denotes the state that satisfies a particular rule.
(2) Construct a product automaton \( \mathbb{A}_P \) by combining the \( \mathbb{A}_E \) and the negation of the LTL formula which indicates situations that violate the LTL rule through accepting states. 
(3) Check and analyze accepting states denoting violations in \( \mathbb{A}_P \).

Modeling the environment as the \( \mathbb{A}_E \) in step 1 entails enumerating all possible environment states, i.e., the states of all spaces and devices, along with the state transitions, i.e., events and actions that update the space and device states.
To mitigate the overhead from extraneous state dimensions when constructing \( \mathbb{A}_E \), EnvGuard extracts relevant states, events, and actions from environment representation for each LTL, and constructs a minimal automaton capturing all relevant environment behaviors for model checking, thereby avoiding unnecessary complexity.

Specifically, EnvGuard initially takes device and space states corresponding to all atomic propositions in the LTL as foundational state dimensions. 
It then queries actions in the environment representation that have effects on the space states within the foundational state dimensions, incorporating the corresponding device states and precondition space states into the state dimensions. 
Next, EnvGuard defines the transitions by the events and actions related to the space or device states in the state dimensions.
Each event transitions the corresponding space state to its associated state value. Action alters the device state and, based on the effect and whether the current environment state meets the precondition, modifies the space state (e.g., the action of turning on the air conditioner changes its device state to "on" and decreases the space temperature).
Finally, EnvGuard constructs the \( \mathbb{A}_E \) by querying the initial values of all state dimensions from the representation and performing a breadth-first search to explore all possible environment states according to the transitions. 
The detailed construction process is illustrated in Algorithm \ref{algo-BA}.

\begin{algorithm}[h]
\SetAlgoLined
\caption{Büchi Automaton \( \mathbb{A}_E \) construction}\label{algo-BA}

\KwIn{LTL formula, environment representation}
\KwOut{Büchi Automaton \( \mathbb{A}_E \)}

Find all \textbf{deviceState} and \textbf{spaceState} in LTL

\For{state in spaceState}{
    effectList $\gets$ findEffect(state, "effect")\;
    \For{effect in effectList}{
    deviceState.append(findDeviceState(effect))\;
    \scalebox{0.99}{spaceState.append(findSpaceState(effect.precondition))\;}
    }
} 
Trans $\gets$ [findEvent(spaceState), findAction(deviceState)]\;

States $\gets$ [deviceState, spaceState]\;

Initialize Büchi Automaton \( \mathbb{A}_E \)  as \textbf{nodeList} and \textbf{transList}\;

initNode $\gets$ getStateValue(States)\;

searchPool $\gets$ [initNode]\;
    
\While{searchPool}{
    node = searchPool.pop()\;
    
    \For{trans in Trans}{
        tempNode $\gets$ applyTrans(trans, node)\;
        
        \If{tempNode not in nodeList}{
            nodeList.append(tempNode)\;
            searchPool.push(tempNode)\;
        }
        transList.append(getTrans(node, trans, tempNode))\;
    }
}

\textbf{Return} Büchi Automaton \( \mathbb{A}_E \) 
\end{algorithm}

To obtain the product automaton \( \mathbb{A}_P \) in step 2, EnvGuard identifies the negation of the LTL property, converting the expected system property into a direct depiction of violations. 
For example, the LTL property G!(Lab.AC.on \( \land \) Lab.Heater.on), which states that the air conditioner and heater should never be turned on simultaneously due to the conflict effect, is negated to F(Lab.AC.on \( \land \) Lab.Heater.on), indicating the situation when both air conditioner and heater are turned on.
Then the negated LTL formula is transformed into a B\"{u}chi Automaton \( \mathbb{A}_N \) where all violations that satisfy the negated LTL formula are marked as accepting states~\cite{Gastin2001FastLT}. 
By combining \( \mathbb{A}_E \) and \( \mathbb{A}_N \) using the Cartesian product, the product automaton \( \mathbb{A}_P \) is derived, which encompasses all potential violations and identifies all violated environment states as accepting states.
Applying the off-the-shelf model checking tool SPOT~\cite{DuretLutz2022FromS2}, EnvGuard accomplishes computations in step 2.

\begin{figure}[htbp]
	\centering
	\includegraphics[width = 0.33\textwidth]{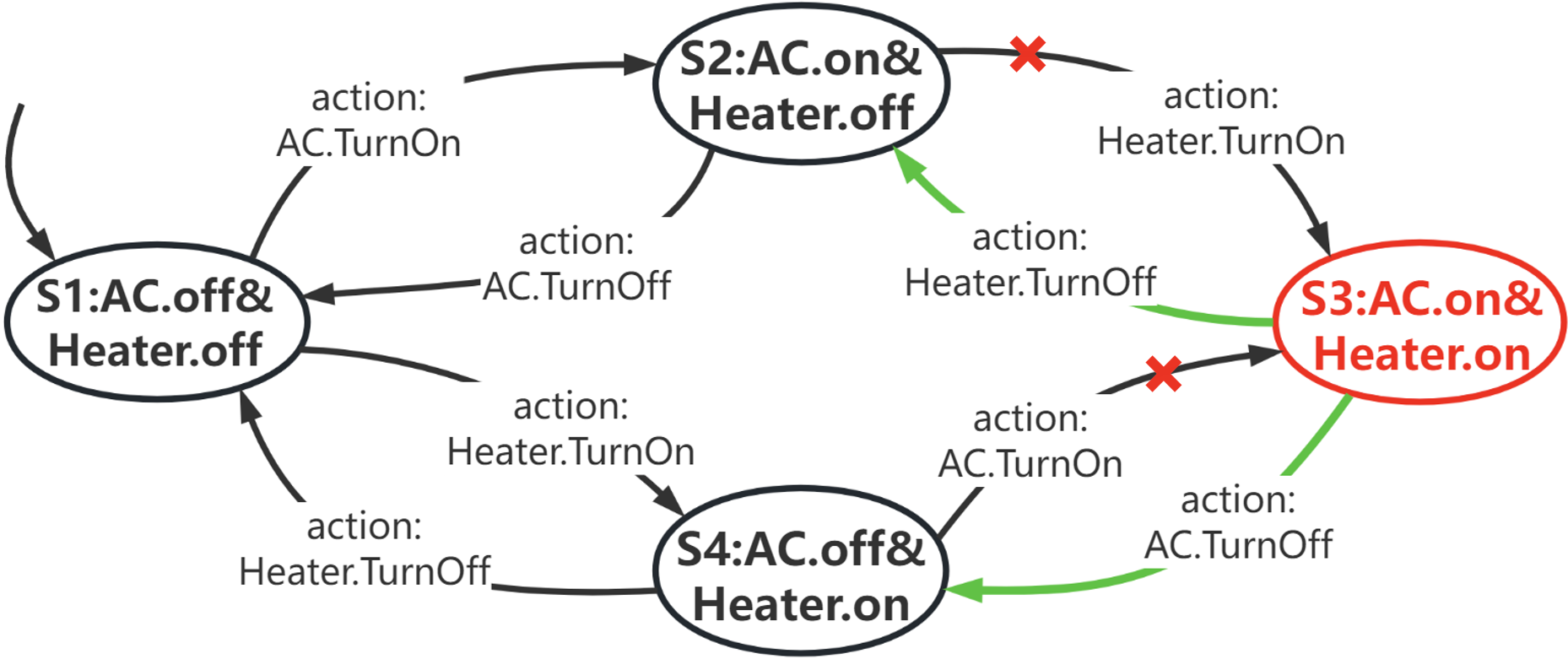}	
    \vspace{-6pt}
	\caption{An example of the product automaton \( \mathbb{A}_P \) where the accepting node S3 indicates the abnormal state and the transitions to the accepting node indicate violations.}
        \Description{An example of the product automaton \( \mathbb{A}_P \) where the accepting node S3 indicates the violated state and the transitions to the accepting node indicate violations.}
    \vspace{-8pt}
	\label{AcceptingNode}
\end{figure}

In step 3, EnvGuard identifies violations by examining transitions in \( \mathbb{A}_P \) leading to accepting states and their prior states, marking them as spatial state violations.
Figure \ref{AcceptingNode} presents an example of a product automaton,
which is constructed to identify the conflict between the air conditioner and the heater. The \( \mathbb{A}_P \) contains all possible environment states and transitions between states starting from the initial state S1. The accepting state S3 is marked red. 
In it, the actions "Heater.TurnOn" in state S2 and "AC.TurnOn" in state S4 are identified and recorded as violations since they transition to the accepting state S3.
EnvGuard records all violations in \( \mathbb{A}_P \) as \textit{<id, Violation\_Trans>}. 
The id indicates the corresponding LTL property of the \( \mathbb{A}_P \).
The Violation\_Trans is a map $\{(RiskState_1:Violation_1), ..., (RiskState_n:Violation_n)\}$ where RiskState $\{(s_1:v_1), ..., (s_n:v_n)\}$ recording the state that a violation may occur,
and Violation $\{t_1, ..., t_n\}$ indicating all the violation transitions (i.e., the violation event or action) under the state.

At runtime, EnvGuard employs the State Checker module to continuously monitor the environment state with event-driven updates and then to check whether the subsequent event or action is a violation under the current environment state. If detected, the State Checker module reports it as a spatial state violation to the Resolution Execution module for further processing.

\subsubsection{Temporal Trace Violation Identification}\label{Temporal-Trace}
To identify temporal trace violations in the environment, EnvGuard develops a MTL parser to translate the MTL formulas into temporal traces of environment states. A temporal trace consists of three parts: (1) A Trigger that illustrates the start of the trace. (2) A Timer that indicates the duration of the trace. (3) A Condition that describes the requisite environment state during the timer range.

\begin{figure}[htbp]
	\centering
	\includegraphics[width = 0.45\textwidth]{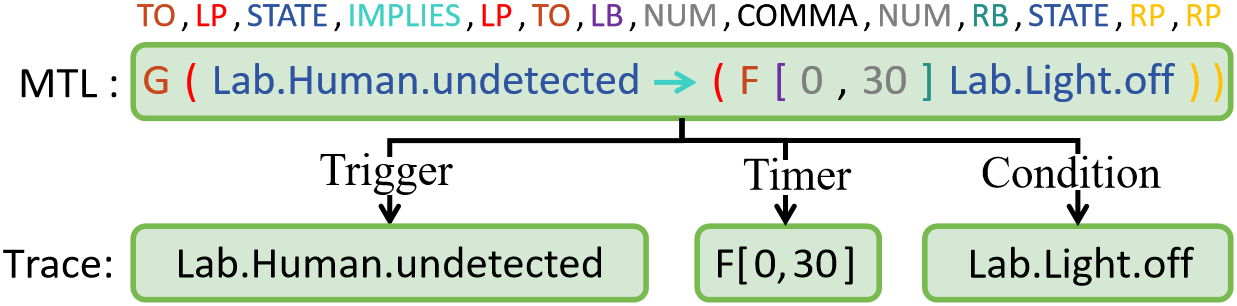}	
    \vspace{-6pt}
	\caption{An example of parsing a MTL formula into a trace.}
        \Description{The MTL formula is parsed into a trigger, a timer, and a condition}
    \vspace{-8pt}
	\label{MTLParser}
\end{figure}

In order to generate traces based on MTL formulas, the MTL parser translates the MTL formula into the Abstract Syntax Tree (AST) for trace conversion.
By defining the lexicon and syntax of the MTL formula according to the property template, the MTL parser performs lexical analysis to tokenize a MTL formula into a sequence of tokens and then conducts syntactic analysis to construct an AST by parsing the token sequence according to the syntax of the MTL formulas. 
Finally, the MTL parser traverses the AST to generate the temporal trace. Figure \ref{MTLParser} demonstrates an example of generating a trace from a MTL formula. 
The MTL formula, which dictates the light should be turned off within 30 seconds after everyone has left the laboratory to conserve energy, is parsed into a trace, which consists of a trigger "G Lab.Human.undetected", a timer "F [0,30]", and a condition "Lab.Light.off". 
By using the lexical and syntax analysis tool PLY~\cite{Mason1992LexY}, the parser implements the process of AST constructing and trace generating. 

\vspace{-3pt}
\vspace{-5pt}

To identify the time when the Condition is violated, we initiate an analysis of all the temporal violation occurrences.
For Timer involving the F temporal operator, a violation occurs if the condition state is never satisfied within the timer.
For Timer involving the G temporal operator, a violation occurs if there's any instance within the timer where the condition state is not satisfied.

\textbf{Observation:} Here emerges a key observation. Regardless of the timer type, the temporal characteristic of a trace is manifested in the timing of the violation trigger, while the underlying cause of the violation is only attributed to the spatial state, independent of the temporal aspect.

This observation allows us to analyze the Condition individually using the same model checking approach introduced in \ref{Spatial-State} to identify the causative behaviors leading to violations. 
EnvGuard generates a product automaton for each trace Condition that captures all environment states violating the condition, and then traverses all accepting states in the product automaton to record the risk states that can transit to a violated state, the transitions denoting violations, and the violated states.
By leveraging the Trigger and Timer to determine the timing of Condition judgment, the temporal trace violation identification is accomplished.

EnvGuard records trace as a tuple \textit{<id, Trigger\_State, Timer, Condition, Violated\_State>}. id indicates the corresponding MTL property of the trace.
Trigger\_State is a map $\{(s_1:v_1), ..., (s_n:v_n)\}$ denoting the state value when the trace is triggered, and Timer specifies the time duration of the trace. Condition is a map $\{(RiskState_1:Violation_1), ..., (RiskState_n:Violation_n)\}$ where each RiskState indicates a state that a violation may occur, and each Violation indicates all the transitions to a violated state from the RiskState. The Violated\_State illustrates all violated states where each state is a map $\{(s_1:v_1), ..., (s_n:v_n)\}$ denoting the state value when the trace is violated.


At runtime, EnvGuard employs the Trace Checker module to monitor the update of environment representation in an event-driven way.
Once the trigger state is met, the Trace Checker starts the timer and then checks whether the condition is satisfied. If the condition is satisfied within the timer, the Trace Checker terminates this tracing. Otherwise, the current state is reported as a temporal trace violation.

\subsubsection{Rationale Discussion} 
The reasons for separate identifying violations of spatial states and temporal traces are mainly threefold:

(1) Spatial states and temporal traces are two distinct features of the environment. By respectively applying LTL formulas, which focus on instantaneous state properties, and MTL formulas that emphasize temporal trace of state, EnvGuard ensures the \textbf{accuracy} of violation identification.

(2) Modeling all the state dimensions with a complex timed automaton to uniformly detect all the LTL and MTL properties can cause state space explosion and add unnecessary complexity. Separately detecting the LTL and MTL properties ensures the \textbf{feasibility} of EnvGuard.

(3) Since EnvGuard uses an event-driven way to monitor the environment state, the overhead caused by polling device states is avoided. It ensures the runtime \textbf{efficiency} for checking LTL and MTL properties separately.

\subsection{Violation Resolving}
We first conduct a user study to investigate the user-preferred resolution action for different types of violations, thereby providing three optional resolving strategies.
Then EnvGuard finds all possible resolution actions (i.e., actions that can transition from a violated state to a non-violated state) for each violation in the offline phase.
Based on the selected resolving strategy for each property, EnvGuard performs a compatible resolution execution during runtime.

\subsubsection{User Study}
We designed a survey questionnaire to investigate the user preference for resolution actions.
The survey contextualized within two typical WoT environments: a smart office and a smart home. For each environment, we initially introduce the specified environment context, including the type of environment, room layout, and deployed devices.
Subsequently, the survey listed several prevalent violations and their available resolution actions, asking participants to select their preferred options. For each environment, eight prevalent spatial state and temporal trace property violation cases are listed. 
For each violation case, we provided the root cause (i.e., the violated property and the action or event leading to the violation) and three or four types of resolution options: 

(1) \textbf{Intercept or revoke the violation action}.

(2) \textbf{Modify the environment state}. Violations may involve multiple space and device states in the environment, and some violations can be resolved by altering the environment state to make them non-violating.

(3) \textbf{Intercept and replace the violation action}. In addition to intercepting the violation action, execute an alternative non-violating action with the same effect.

(4) Any of the above is acceptable. 

To encourage diversity, the survey also provided an additional option for participants to propose other resolution actions that are not listed in the survey. The full survey questionnaire is available\cite{EnvGuard}.

We recruited 89 students from diverse majors within the campus to complete the survey questionnaire, including 56 males and 33 females. Thirty-nine of them had experience in using WoT applications. Their educational backgrounds varied  from undergraduate to graduate students. During the user study, we ensured participants' informed consent and privacy protection.

According to the results of the survey, a significant majority of the participants showed personal preferences for resolution actions. For spatial state property violations, 54.9\% participants selected the intercept and replace options, 25.0\% participants selected to modify the environment state, and 18.1\% participants selected the intercept or revoke options.
For temporal trace property violations, 65.9\% participants selected the intercept or revoke options, 28.4\% participants selected to modify the environment state, and 2.8\% participants selected the intercept and replace options.

From the participant feedback and our analysis, we found that spatial state violations often result from users' immediate intentions with operational oversights. Therefore, preserving user intent during resolution is crucial. Temporal trace violations mainly stem from overlooking a device's prolonged impact on the environment, necessitating timely device state modifications. 
Furthermore, some participants suggested adding a message notification option to notify users for manually handling the violations.


\vspace{-5pt}
\subsubsection{Violation Resolution}\label{SubSec:Repair-Action}
To find resolution actions for both spatial state and temporal trace violations, EnvGuard traverses all accepting states in the corresponding product automaton for each LTL and MTL property, identifying action-type transitions leading from accepting to non-accepting states. 
As the example given in Figure \ref{AcceptingNode}, the action "Heater.TurnOff" and "AC.TurnOff" in state S3 are identified as the resolution actions, as they can transition the accepting state S3 to non-accepting state S2 or S4. 

For each LTL and MTL property, EnvGuard records all resolution actions as \textit{<id, Resolution>} where id indicates the corresponding LTL or MTL property, and Resolution records all executable resolution actions for each accepting state in form of \textit{<Violation\_Transition, State, Action>}. Violation\_Transition is a list of transitions leading to the accepting state, State specifies the accepting state, and Action records actions that can transition the accepting state to a non-accepting state.

To satisfy user preferences for resolving property violations, EnvGuard offers three resolving strategies mentioned before, along with an additional message notification option, and further supports users to select their preferred  strategy when constructing properties.
At runtime, EnvGuard employs the Resolution Execution module to select and execute the optimal resolution action for each violation.
For example, if the window is opened during rain and the intercept and replace strategy is selected, the Resolution Execution will close the window (intercept the violation) and activate both the humidifier and air purifier (replace the window open action with alternatives that have humidification and air purification effects). 
If no resolution action is available, the Resolution Execution notifies the user for manual intervention.

To prevent the resolution actions from introducing new violations, the Resolution Execution assesses each action before execution. If either the State Checker or the Trace Checker identifies a resolution action as non-compliant in the current environment state, the Resolution Execution refrains from executing it and selects the next available action for assessment.  
If multiple properties are violated by one event or action, EnvGuard generates and executes a resolution action for each property separately.
To optimize runtime efficiency, the State Checker and the Trace Checker proactively notify the Resolution Execution of the non-compatible actions under current environment state, which are then removed from the resolution action list, thereby avoiding the extra time consumed by queries in resolving conflicts between properties.

\vspace{-5pt}
\section{Case Study}
We evaluate EnvGuard in two representative real-world WoT environments to answer the following research questions:

\textbf{RQ1 (Usability):} How well can end users specify environment properties by using the provided template-based GUI tool to meet the safety and security requirements of real-world scenarios?

\textbf{RQ2 (Feasibility):} Whether the EnvGuard can identify violations accurately and generate
resolution actions aligning with user intention?

\textbf{RQ3 (Efficiency):} How efficiently does the violation identification and resolving perform during runtime? 

\subsection{Experiment Settings}
\begin{figure}[h]
	\centering
	\includegraphics[width = 0.45\textwidth]{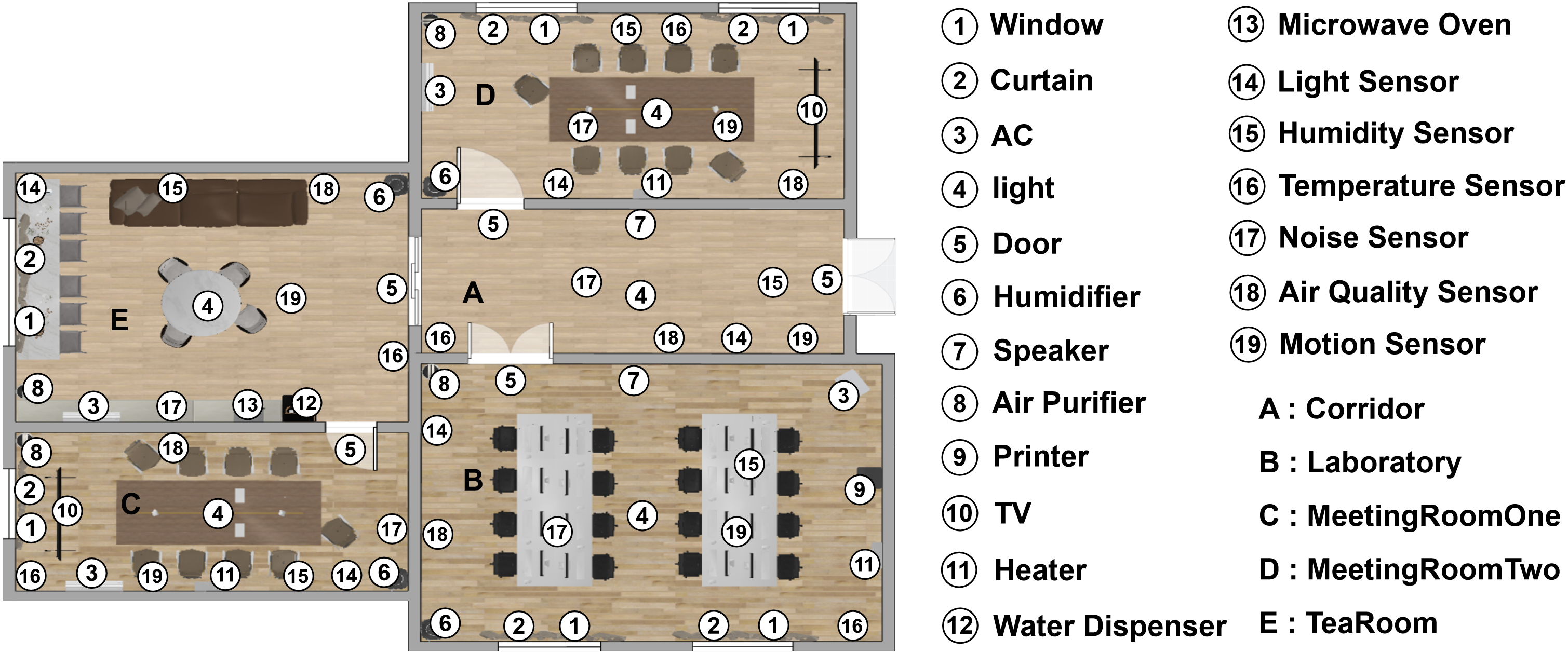}
        \vspace{-8pt}
	\caption{Smart office environment with deployed devices}
        \Description{Various devices are deployed in the office environment}
    \label{Office}
\end{figure}

\begin{figure}[h]
	\centering
	\includegraphics[width = 0.45\textwidth]{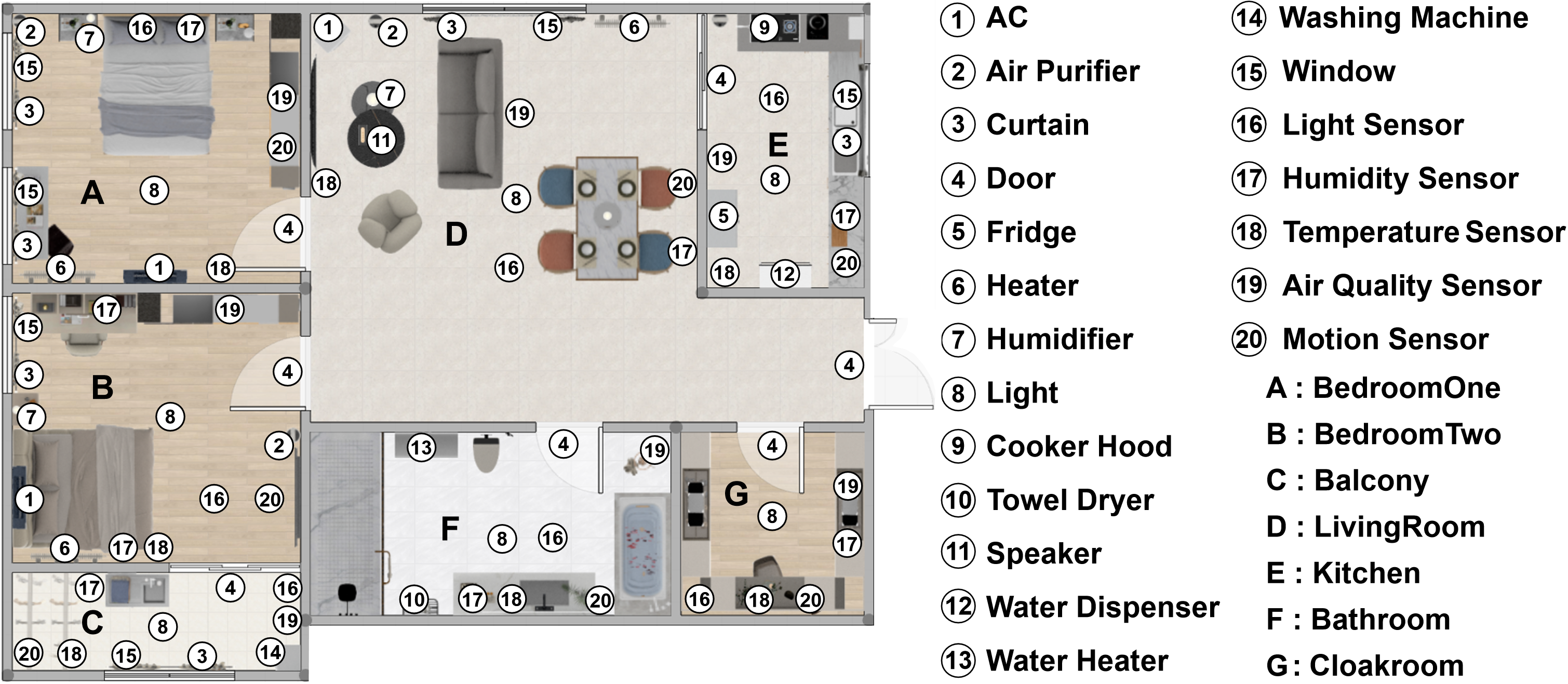}
        \vspace{-8pt}
	\caption{Smart home environment with deployed devices}
        \Description{Various devices are deployed in the home environment}
    \label{Home}
\end{figure}

We evaluate EnvGuard in two real-world WoT environments including a smart office and a smart home, where multiple individuals work or live daily.
The spatial layout and the deployed devices of the office and home environments are illustrated in Figure~\ref{Office} and Figure~\ref{Home}, respectively. 
In the office environment, there are 44 actuators of 13 different types and 30 sensors of 6 different types deployed. 
Ten applications are deployed in the office to manage humidity levels, control TV switches, operate air conditioner, automatically prepare reserved meeting rooms, and manage staff check-ins.
In the home environment, there are 43 actuators of 15 different types and 35 sensors of 5 different types deployed. 
Eight home automation applications are also deployed for bath water preparation and towel drying, air quality management, morning wake-up, laundry progress regulation, and lighting control. More details of the environment setting can be seen online\cite{EnvGuard}

\vspace{-8pt}
\vspace{-15pt}

\vspace{-8pt}

To make the evaluation more targeted, we obtained ten expected safety and security property requirements from interviews with individuals who work or live there daily for each environment, as shown in Table~\ref{office-property} and Table~\ref{home-property}.
Among the requirements, ten spacial state properties (\#1 to \#5 and \#11 to \#15) and ten temporal trace properties (\#6 to \#10 and \#16 to \#20) are included.

\vspace{-3pt}
\begin{table}[H]\small
  \caption{Smart office property requirements}
  \vspace{-8pt}
  \label{office-property}
  \renewcommand\arraystretch{1.1}
  \resizebox{\linewidth}{!}{
  \begin{tabular}{c p{7.5cm}}
      \toprule
      Ord. & Requirement Description  \\
      \midrule
      \#1 & To prevent temperature regulation deterioration, simultaneous operation of a warming device and a cooling device is prohibited. \\
      \#2 & To ensure sufficient brightness, devices with a brightness-increasing effect must be activated when someone is detected. \\   
      \#3 & To prevent rain from soaking contents, windows should not be opened when it is raining. \\ 
      \#4 & To use electrical appliances safely, the microwave oven should not be used unsupervised. \\ 
      \#5 & To use electrical appliances safely, the water dispenser should not be used unsupervised. \\ 
      \#6 & To prevent excessive humidity, the device with a humidifying effect should not operate continuously for more than 60 minutes. \\ 
      \#7 & To prevent the health risks of low air quality, air purification must be executed in occupied room within 10 minutes of low air quality. \\ 
      \#8 & To prevent wasting electricity, lights must be switched off within 30 seconds when nobody is detected. \\ 
      \#9 & To prevent the water dispenser tank from drying out, the water dispenser cannot operate continuously for more than 2 minutes. \\ 
      \#10 & To prevent excessive noise caused by continuous speaker operation, the speaker cannot operate continuously for more than 1 minute. \\ 
      
\bottomrule
\end{tabular}
}
\end{table}

\vspace{-8pt}
\begin{table}[H]\small
  \caption{Smart home property requirements}
  \vspace{-8pt}
  \label{home-property}
  \renewcommand\arraystretch{1.1}
  \resizebox{\linewidth}{!}{
  \begin{tabular}{c p{7.5cm}}
      \toprule
      Ord. & Requirement Description  \\
      \midrule
      \#11 & To prevent temperature regulation deterioration, simultaneous operation of a warming device and a cooling device is prohibited. \\
      \#12 & To ensure a suitable temperature, the operation of a cooling device is prohibited when someone is detected and the temperature is low. \\   
      \#13 & To ensure the privacy of the home, windows should always be closed when the home is empty. \\ 
      \#14 & To prevent rain from soaking contents, windows should not be opened when it is raining. \\ 
      \#15 & To ensure the privacy of the home, curtains should only be opened only when someone is detected. \\ 
      \#16 & To prevent excessive humidity, the device with a humidifying effect should not operate continuously for more than 60 minutes. \\ 
      \#17 & To prevent the health effects of low air quality, air-improving devices must be activated within 5 minutes of low air quality. \\ 
      \#18 & For security, the door must be closed within one minute of everyone leaving the home. \\ 
      \#19 & To prevent wasting electricity, the fridge should be open for a maximum of 5 minutes. \\ 
      \#20 & To use electrical appliances safely, the towel dryer should be open for a maximum of 60 minutes. \\ 
      
\bottomrule
\end{tabular}
}
\end{table}

\vspace{-17pt}
\subsection{Implementation and Initialization}
We implemented EnvGuard as a prototype, which included a GUI property configuration tool using JavaScript, an environment representation service using Python, offline analysis services using Python and C++, and a runtime engine using Java.
The prototype is deployed in a Lenovo laptop with Intel Core i7-13650HX CPU and 64GB memory.
To initiate EnvGuard, 1518 lines of configuration for 74 devices and 92 lines of configuration for 6 spaces (including outdoor area) within the office setting, as well as 1550 lines of configuration for 78 devices and 104 lines of configuration for 8 spaces (including outdoor area) within the home setting, are supplied to construct environment representations (available online\cite{EnvGuard}). 
\vspace{-6pt}

\subsection{Usability}
To answer RQ1, we conducted a second user study requiring participants to specify the properties for each environment through the property configuration tool. We invited 21 students (including 7 females) as participants who were familiar with the office and home environment. These students majored in software engineering with different programming experiences. Among the participants, eight of them have experience using WoT platforms and developing WoT applications, while the others do not.

Before the study, we provided a brief tutorial to introduce the key concepts (e.g., event, action, effect, property, etc.) and the usage of the tool. After a demonstration example, participants are further asked to complete several attention-check questions to ensure their understanding. 
Then, participants start to specify properties in sequential order based on the requirement descriptions.
The constructed properties can be seen in our replication package\cite{EnvGuard}. During the process, we record the time consumption for each property specifying.

Table \ref{usability-table} summarizes the average time consumption of the two groups (Group A contains 8 participants with WoT experience and Group B contains the rest 13) for each property configuration. 
From the statistics, we observe that the time spent by participants with experience with WoT experience is not significantly less than non-expert participants. 
This is because our template-based property description method does not require users to have expertise.
Users only need to intuitively describe the expected device behaviors in the environment.
Additionally, we find that configuring temporal trace properties is more time-consuming than spatial state properties. 
This is because participants need to spend more time thinking about the timing of the violations.
Furthermore, we find that the time consumption for property configuration gradually decreased during the process, and the time required for constructing similar properties (e.g., \#4 "water dispenser should always be used under supervision" and \#5 "microwave oven should always be used under supervision") significantly reduced.
This indicates that our proposed method exhibits great learnability.

\vspace{-8pt}
\begin{table}[htbp]\footnotesize
  \caption{Average consumed time (minutes) of two groups to accomplish environment property configuration}
  \vspace{-10pt}
  \label{usability-table}
  \renewcommand\arraystretch{0.9}
  \setlength{\aboverulesep}{0.8pt}
  \setlength{\belowrulesep}{1.1pt}
  \resizebox{\linewidth}{!}{
  \begin{tabular}{@{}ccccccccccc@{}}
  \toprule
  \multicolumn{1}{c|}{}  & \#1  & \#2  & \#3  & \#4  & \#5  & \#6  & \#7  & \#8  & \#9  & \#10 \ \ \ \ \\ 
  \midrule 
  \multicolumn{1}{c|}{\textit{A}} & 7.3  & 5.8  & 2.6  & 2.9  & 1.2  & 7.5  & 6.3  & 4.9  & 4.6  & 4.5 \ \ \ \ \\
  \multicolumn{1}{c|}{\textit{B}} & 8.2  & 6.2  & 3.4  & 3.1  & 1.1  & 7.9  & 6.2 & 5.2  & 4.9  & 4.6 \ \ \ \ \\  \midrule
  \multicolumn{1}{c|}{}  & \#11  & \#12  & \#13  & \#14  & \#15  & \#16  & \#17  & \#18  & \#19  & \#20 \ \ \ \ \\  \midrule
  \multicolumn{1}{c|}{\textit{A}} & 4.7  & 5.3  & 4.5  & 2.1  & 2.9  & 4.1  & 3.5  & 1.7  & 1.5  & 1.2 \ \ \ \ \\
  \multicolumn{1}{c|}{\textit{B}} & 4.1  & 5.1  & 4.4  & 1.8  & 2.8  & 3.9  & 3.2  & 1.6  & 1.7  & 1.1 \ \ \ \ \\
  \bottomrule
  \end{tabular}
  }
  \vspace{-8pt}
\end{table}

On the other hand, we invited four experts (including two system developers) to check whether the constructed 420 properties are consistent with the requirements. Through the joint inspection, 29 incorrect properties were identified, including 21 temporal trace properties and 8 spatial state properties. By further interviewing, we find that errors mainly derive from misunderstanding the timing of the temporal trace violations. Taking \#7 "air purification must be executed in occupied room within 10 minutes of low air quality" as an example, four participants mistakenly only consider the low air quality as the activation of the trigger, but ignoring the condition of occupants in the room.

In summary, the results of the user study confirm the usability of the provided property description tool. 
Participants generally agree that describing environment properties through the effects of device services reduced the complexity.
Some participants with WoT development experience note that current property description templates are not sufficient when handling more complicated properties.
They suggest that the tool can be improved by providing a more flexible way to describe properties for professional developers.
In fact, EnvGuard is not restricted to identifying and resolving specific types of violations.
Experts can directly specify environment properties using LTL and MTL formulas as an alternative to the template-based tool, then proceed with EnvGuard's subsequent processes.
Additionally, property description templates can also be extended for specific scenario needs.

\subsection{Feasibility}
To answer RQ2, we apply EnvGuard for a 14-day continuous evaluation in both office and home environments to test if EnvGuard can identify violations accurately and generate resolution actions. 
For comparison, we also implemented and deployed the previous system~\cite{Wang2019ChartingTA, IoTSafe, AutoTap, IOTMEDIATOR} to analyze WoT application and environment information to identify violations.

EnvGuard starts with the offline analysis of the environment properties for each environment, identifying potential violations as well as resolution actions.
The average time cost to analyze each property is approximately 2 minutes, ranging from 0.9 seconds (\#4) to 5.2 minutes (\#11).
After the offline analysis, EnvGuard monitors and processes events and actions from the initial environment context, updates space and device states accordingly, and checks for property violations. For spatial state properties, we select the intercept and replace strategy to generate resolution actions. For temporal trace properties, we select the intercept or revoke strategy for violation resolving.

During the evaluation, we sequentially recorded all the events and actions and their timestamps  from the initial state for each environment.
For each action log, we additionally recorded its source (i.e., whether it is invoked by the application or controlled offline by the user).
Then we invited six experts with WoT development experience to independently analyze and label the events and actions with violated properties (Fleiss Kappa = 0.68) and resolve discrepancies through discussion to obtain the ground truth.
In the office environment, an average of about 1300 events and actions are recorded each day, with a total of 310 violations labeled in the collected data.
In the home environment, an average of about 310 events and actions are recorded each day, with a total of 268 violations labeled.
Due to the lack of publicly available dataset that includes device, environment state changes, WoT application log, and user activities for detecting violations in environments, we construct the collected and labeled environment data as a first dataset to encourage further research. Detailed information about the dataset is available online~\cite{EnvGuard}.

The results for violation identification in the office environment and the home environment are separately presented in the Table~\ref{v-office} and Table~\ref{v-home}, where EnvGuard successfully identifies all violations. 
We further find that in real-world environments, the majority of violations originate from user behavior and environment state change, with only a minority being caused by applications. Therefore, methods that lack an analysis of the interplay between device and environment (iRuler, IoTSafe, AutoTap) are limited to identify all violations. 
Meanwhile, methods that exclusively identify specific types of violations (iRuler, IoTSafe, IOTMEDIATOR) fail to detect customized properties. 
Furthermore, conducting checks solely on spatial states (iRuler, IoTSafe, IOTMEDIATOR) results in the inability to detect temporal property violations.

\vspace{-10pt}
\begin{table}[!htb]
  \caption{Violation in office environment}
  \label{v-office}
  \renewcommand\arraystretch{1.15}
  \vspace{-10pt}
  \resizebox{\linewidth}{!}{
    \begin{tabular}{|c|cc|cc|}
    \hline
    \multirow{2}{*}{\textbf{}} & \multicolumn{2}{c|}{\textbf{Spatial State}}    & \multicolumn{2}{c|}{\textbf{Temporal Trace}}   \\ \cline{2-5} 
    & \multicolumn{1}{c|}{Application} & Environment & \multicolumn{1}{c|}{Application} & Environment \\ \hline
    iRuler~\cite{Wang2019ChartingTA}    & \multicolumn{1}{c|}{0/0}    & 0/156    & \multicolumn{1}{c|}{0/21}      & 0/133       \\ \hline
    IoTSafe~\cite{IoTSafe}  & \multicolumn{1}{c|}{0/0}    & 0/156    & \multicolumn{1}{c|}{0/21}      & 0/133       \\ \hline
    IoTMEDIATOR~\cite{IOTMEDIATOR}  & \multicolumn{1}{c|}{0/0}   & 88/156   & \multicolumn{1}{c|}{0/21}   & 0/133       \\ \hline
    AutoTap~\cite{AutoTap}      & \multicolumn{1}{c|}{0/0}   & 88/156   & \multicolumn{1}{c|}{12/21}  & 109/133     \\ \hline
    \textbf{EnvGuard}     & \multicolumn{1}{c|}{\textbf{0/0}}   & \textbf{156/156}  & \multicolumn{1}{c|}{\textbf{21/21}}  & \textbf{133/133}     \\ \hline
\end{tabular}
}
\end{table}

\vspace{-15pt}
\begin{table}[!htb]
  \caption{Violation in home environment}
  \label{v-home}
  \renewcommand\arraystretch{1.15}
  \vspace{-10pt}
  \resizebox{\linewidth}{!}{
    \begin{tabular}{|c|cc|cc|}
    \hline
    \multirow{2}{*}{\textbf{}} & \multicolumn{2}{c|}{\textbf{Spatial State}}    & \multicolumn{2}{c|}{\textbf{Temporal Trace}}   \\ \cline{2-5} 
    & \multicolumn{1}{c|}{Application} & Environment & \multicolumn{1}{c|}{Application} & Environment \\ \hline
    iRuler~\cite{Wang2019ChartingTA}    & \multicolumn{1}{c|}{2/2}    & 0/191    & \multicolumn{1}{c|}{0/10}      & 0/65       \\ \hline
    IoTSafe~\cite{IoTSafe}  & \multicolumn{1}{c|}{2/2}    & 0/191    & \multicolumn{1}{c|}{0/10}      & 0/65       \\ \hline
    IoTMEDIATOR~\cite{IOTMEDIATOR}  & \multicolumn{1}{c|}{2/2}   & 113/191   & \multicolumn{1}{c|}{0/10}   & 0/65       \\ \hline
    AutoTap~\cite{AutoTap}      & \multicolumn{1}{c|}{2/2}   & 113/191   & \multicolumn{1}{c|}{8/10}  & 45/65     \\ \hline
    \textbf{EnvGuard}     & \multicolumn{1}{c|}{\textbf{2/2}}   & \textbf{191/191}  & \multicolumn{1}{c|}{\textbf{10/10}}  & \textbf{65/65}     \\ \hline
\end{tabular}
}
\end{table}

For violation resolving, EnvGuard generates resolution actions for all violations with one exception \#7 "In occupied areas poor air quality should be improved within 10 minutes". 
After analyzing the environment context when the violation occurred, we found that the violation is caused by an event AirQuality\_Poor. Since no device with the AirQuality\_Improve effect is deployed in the corridor, the only transition out of the violated state is the uncontrollable AirQuality\_Moderate event, making active resolution of this violation unfeasible. 
This situation can be avoided by alerting on unresolved violations during the offline phase and deploying additional devices with the required effect in advance.

To evaluate whether the generated resolution actions align with users expectation, we conducted a third user study. 
For each property, we randomly selected a violation case and the generated resolution action as a scenario question, inviting the 21 participants to evaluate whether the resolution action aligns with their expectations.
Out of 420 responses, 388 favored the generated resolution actions. Through further feedback, the reason for disagreeing with the resolution actions are mainly because they prefer other resolving strategies, which can be handled by altering the strategy.

Overall, EnvGuard demonstrates superior violation detection performance compared to existing state-of-the-art approaches, and the generated violation resolution actions basically align well with user intention. We believe EnvGuard provides a feasible way for violation identification and resolving.

\subsection{Efficiency}
\begin{figure}[htbp]
	\centering
	\begin{minipage}[c]{0.23\textwidth}
		\centering
		\includegraphics[width=\textwidth]{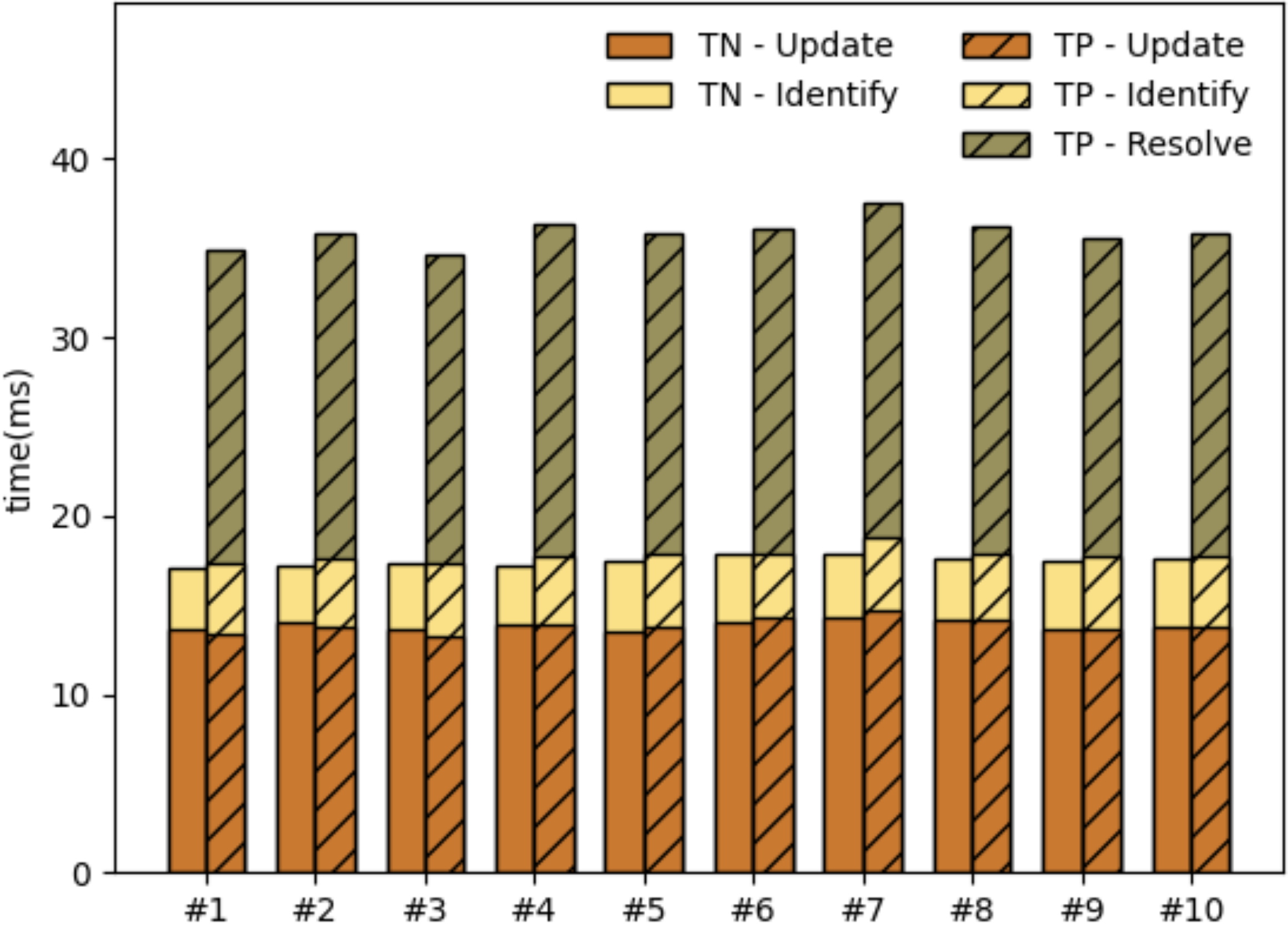}
		\subcaption{office environment}
	\end{minipage} 
	\begin{minipage}[c]{0.23\textwidth}
		\centering
		\includegraphics[width=\textwidth]{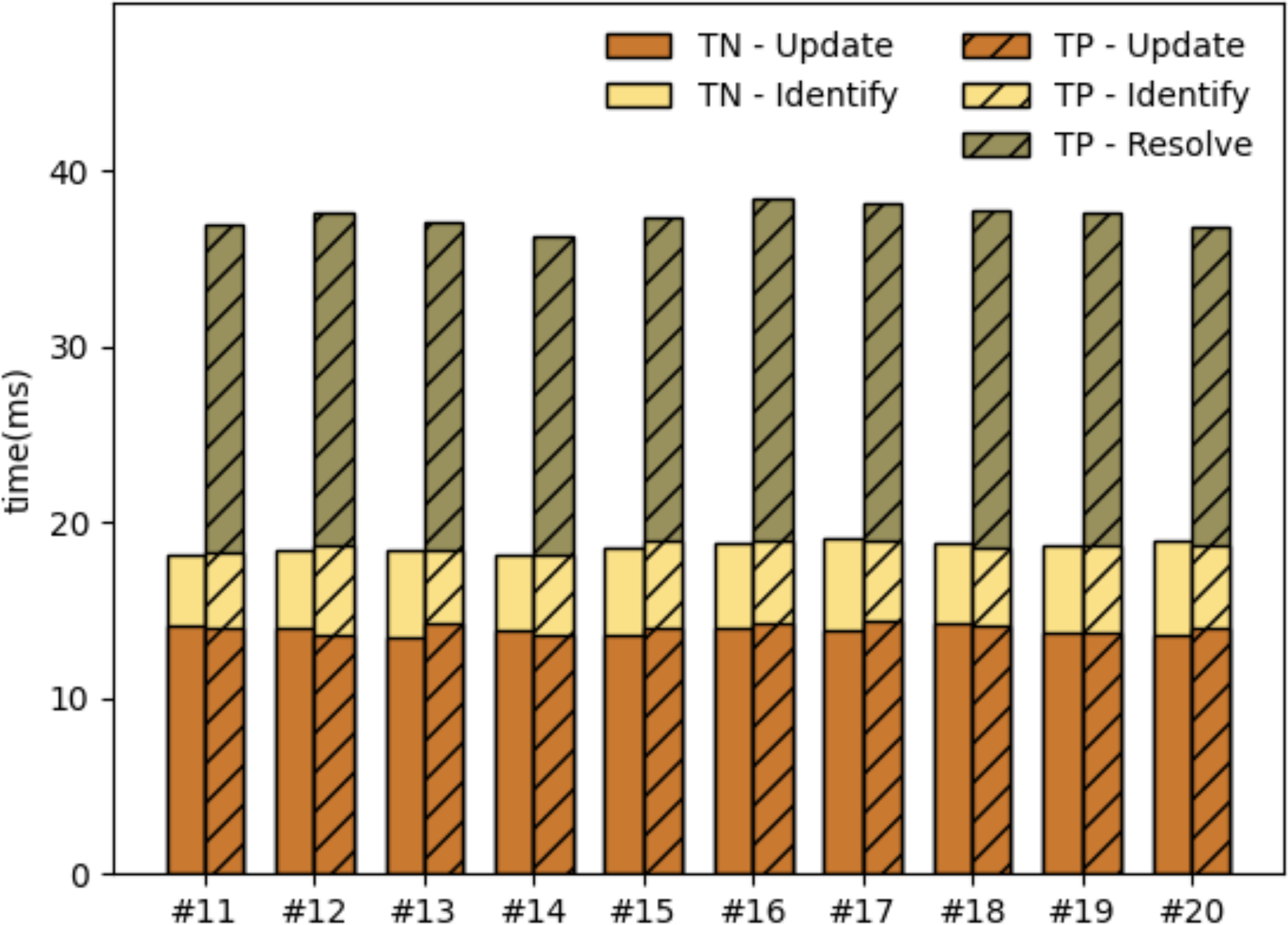}
		\subcaption{home environment}
	\end{minipage} 
        \vspace{-10pt}
	\caption{Runtime execution efficiency}
        \Description{The statistical results of runtime execution efficiency for different properties}
   	\label{Efficiency}
\end{figure}

To answer RQ3, we count and depict the average time consumption on runtime execution that are categorized as TP (true positive indicates a violation is identified and resolved) and TN (true negative indicates a non-violating event or action is correctly identified). 
The results are shown in Figure~\ref{Efficiency}. The time consumption is defined as the interval between monitoring a new event or action and completing of the violation identification and resolution process. 
Furthermore, we divide the time consumption into three parts: the time spent on updating the environment representation, the time spent on the violation identification, and the time spent on the violation resolution. 

According to the results, the time consumption for violation identification and resolution varies little between different properties, which is basically around 25 milliseconds. In addition, the case of TN has a shorter time consumption than TP because no resolution is required.

\begin{figure}[htbp]
	\centering
	\begin{minipage}[c]{0.23\textwidth}
		\centering
		\includegraphics[width=\textwidth]{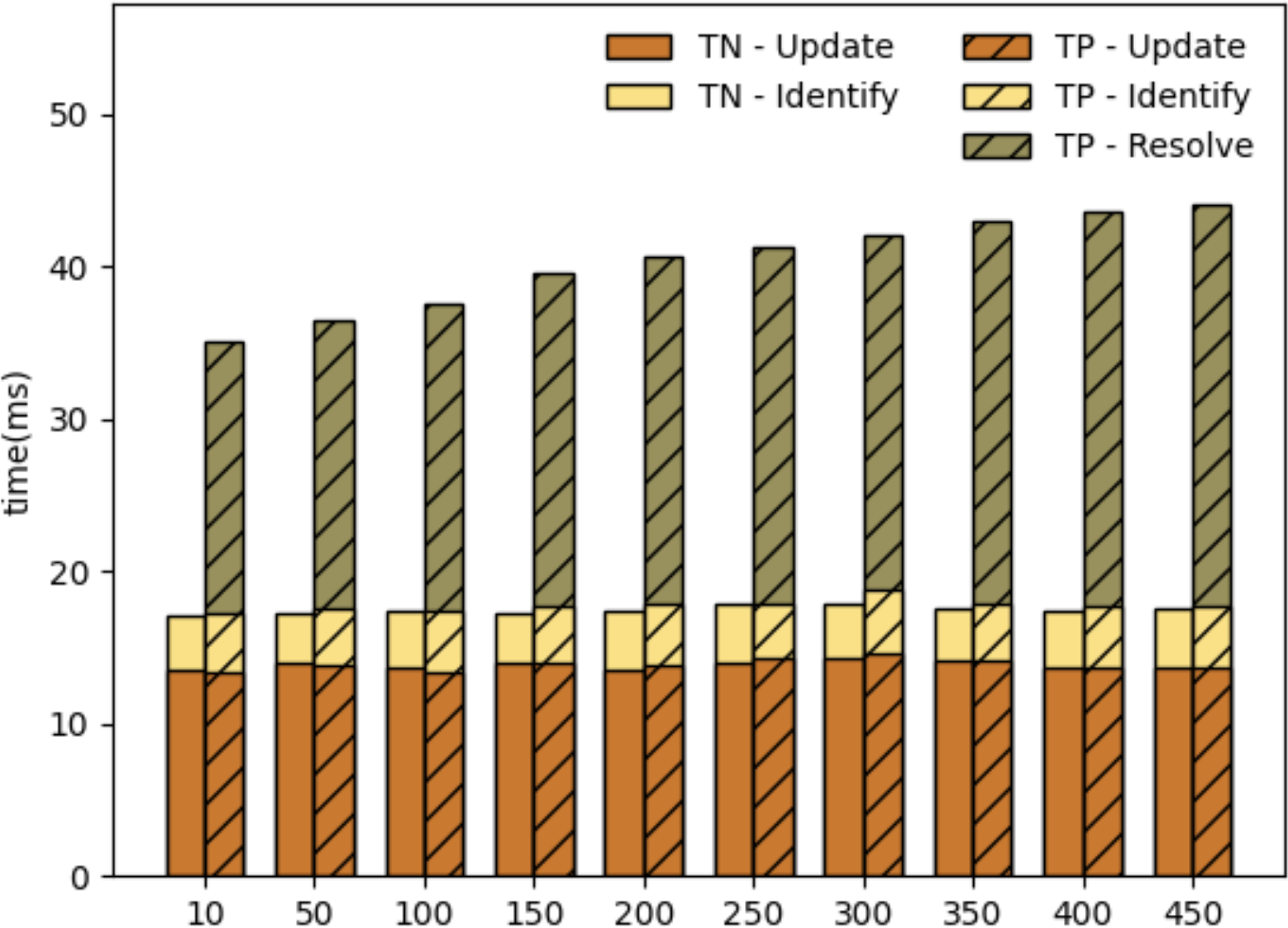}
		\subcaption{office environment}
	\end{minipage} 
	\begin{minipage}[c]{0.23\textwidth}
		\centering
		\includegraphics[width=\textwidth]{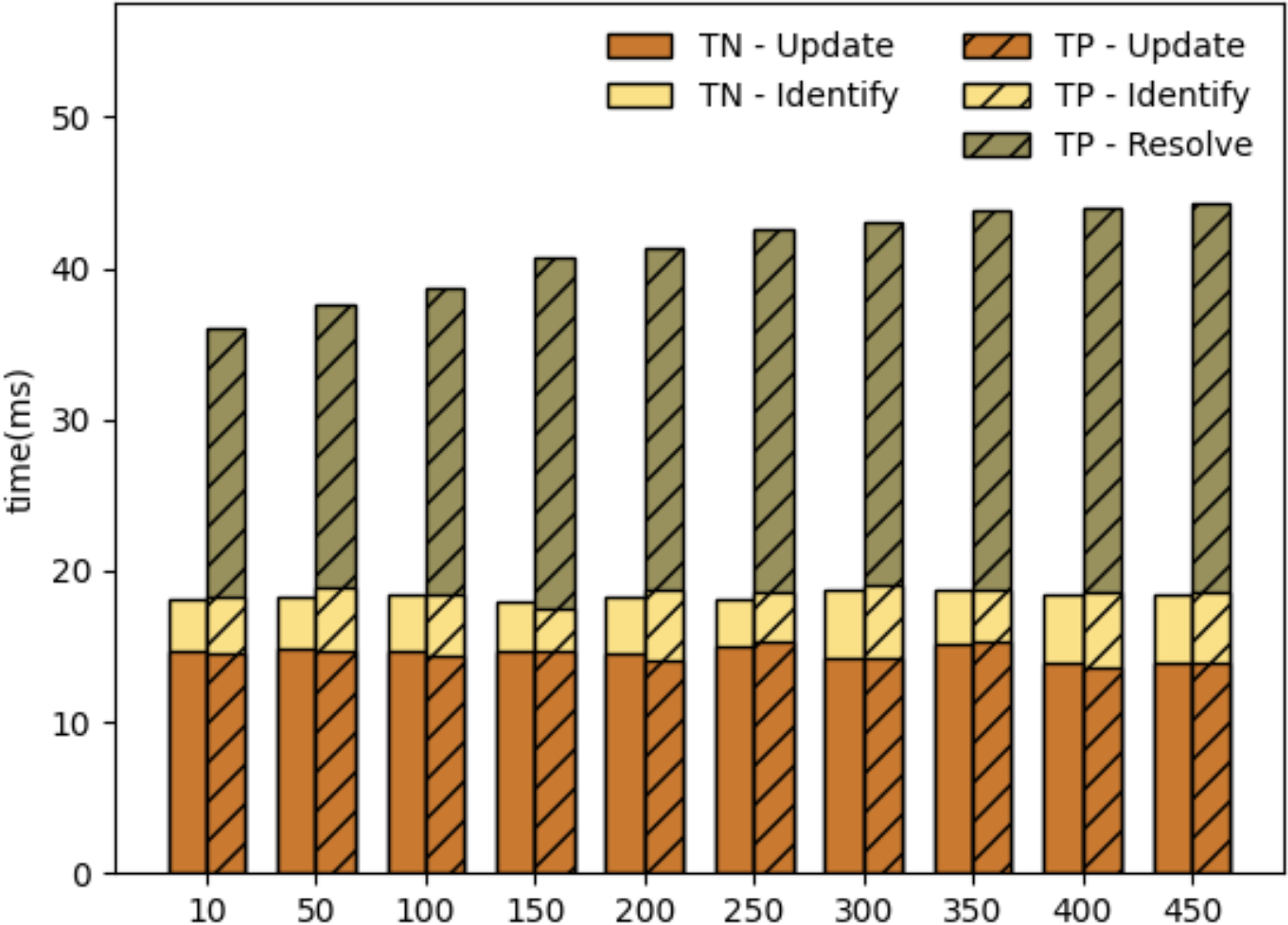}
		\subcaption{home environment}
	\end{minipage} 
        \vspace{-10pt}
	\caption{Overall efficiency}
   	\label{Scalability}
\end{figure}

To further evaluate whether the runtime efficiency is affected by the number of properties, 
we use our construct dataset to simulate real-world environment, and increase the number of properties by duplicating them and counting the overall average time consumption with different numbers of properties, as shown in Figure~\ref{Scalability}.

\vspace{-5pt}

As property count rises, the time cost for environment updates and violation checks remains consistent. This is because updating the representation only involves modifying the environment state value in the graph database, independent of the properties. Moreover, each property is verified independently, without interference from others. The time consumption for violation resolution increases slightly with the number of properties since the concurrent queries caused by simultaneous violations in the duplicated properties slightly affect the time performance.

In real-world scenarios, time delays in WoT systems primarily arise from network communication overhead~\cite{Sun2023SCTAP}. In conventional WoT platforms such as IFTTT\cite{IFTTT}, WoT application responses can experience delays ranging from 2 to 60 minutes\cite{Dong2022RealTimeEO}. Therefore, we believe that the runtime execution of EnvGuard does not cause perceptible time delays, which is acceptable and basically guarantees the real-time efficiency.

\section{Threats To Validity}


First, EnvGuard requires accurate device functionality descriptions for property verification and violation resolution, entailing manual characterization from developers or vendors, potentially incurring additional costs.
Considering that the initialization is one-time for each environment, we believe its impact is minor.
Additionally, automatically generating device descriptions from historical environment data presents a potential solution.

Another threat of EnvGuard is its discrete environment modeling, neglecting the continuous nature of space and devices. This can be refined by using timed modeling for more granular analysis.


In violation resolution, EnvGuard selects single-step actions, augmenting with complementary actions to uphold user intention. Multi-step action collaboration for violation resolution requires deeper analysis tailored to specific scenario characteristics.


Additionally, the environment properties constructed by the user through the provided tools may contain semantic errors, which also pose threats to correctly identifying and resolving violations. Providing more auxiliary information could alleviate this problem.


\section{Related Work}
We outline the related work from the following four aspects.

\textbf{Property Definition.}
Many researches predefined safety and security properties for specific WoT scenario requirements.
In smart home scenarios, some research predefines security violations caused by cyclic invocations between multiple applications in the environment\cite{IOTMEDIATOR, IoTSafe}. Other studies investigate the conflicting effects arising from interactions between different devices\cite{A3ID, Alrawi2019SoKSE, Fernandes2016SecurityAO}.
In industry IoT scenarios, researches have also examined device violations resulting from attacks on device authority~\cite{Tange2019TowardsAS, Pal2021AnalysisOS} and device execution delays caused by network communication blockages~\cite{Zhou2018AnomalyDM, Chaudhari2023AnAO}. 
This approach of predefining properties for specific types of violations fails to satisfy the need for customization.
EnvGuard provide multiple property description template and a GUI tool to assist users in customizing spatio-temporal environment properties.

\textbf{Violation Identification.}
Existing works primarily focus on application-level violations, with semantic-based \cite{A3ID, Tian2017SmartAuthUA} and formalized \cite{Yagita2015AnAC, Croft2015SystematicallyET, Liang2016SystematicallyDI} methods widely adopted. Some studies like AutoTap \cite{AutoTap} and MenShen\cite{Bu2018SystematicallyET} further address device-level violations by analyzing correlations between devices. 
Nevertheless, these studies neglect the analysis of interplay between devices and the environment context, and furthermore, lack the recognition of temporal trace violations.
EnvGuard utilizes model checking to offer a comprehensive guarantee from an environment-centric perspective and complementing previous research.

\textbf{Violation Resolution.}
Existing methods mainly restrict applications during violations \cite{Celik2019IoTGuardDE, Munir2014DepSysDA, Jia2017ContexloTTP} or apply static fixes\cite{Yu2021TAPInspectorSA, Liang2015SIFTBA, Wang2019ChartingTA}. 
EnvGuard provides selectable resolving strategies for different types of properties to ensure alignment with user intentions.

\textbf{Dataset.}
Several IFTTT (If This Then That) datasets \cite{Xu2019PrivacyLI, Yu2021AnalysisOI, Ur2016TriggerActionPI, Kalantari2022ListingTI, Mi2017AnEC} have been built by collecting TAP application, and are widely used for application violation studies. 
At present, there is a lack of datasets for detecting environment-level violations that include data on devices, environment state changes, WoT application logs, and user activities. Our dataset is designed to addresses this gap.

\section{Conclusion}

In WoT environment populated with numerous devices and applications, guaranteeing that complex device behaviors meet the desired environment properties is essential.
In this work, we present EnvGuard, which models the interplay between the context-sensitive device services and the environment context to help users customize safety and security properties from an environment-centric perspective, then further identify the property violations and execute resolving action consistent with user intent.
The evaluation demonstrates the superiority of EnvGuard compared to previous state-of-the-art work, and confirms its usability, feasibility, and runtime efficiency.

\clearpage


\bibliographystyle{ACM-Reference-Format}
\bibliography{reference}



\end{CJK}
\end{document}